# Vector-substrate-based Josephson junctions

Yu-Jung Wu[1], Martin Hack[2], Katja Wurster[2], Simon Koch[2], Reinhold Kleiner[2], Dieter Koelle[2], Jochen Mannhart[1] & Varun Harbola[1]*

[1]Max Planck Institute for Solid State Research

Heisenbergstrasse 1, 70569 Stuttgart, Germany

[2]Physikalisches Institut, Center for Quantum Science (CQ) and LISA+

Universität Tübingen, auf der Morgenstelle 14, 72076 Tübingen, Germany

We present a way to fabricate bicrystal Josephson junctions of high-$T_c$ cuprate superconductors that does not require bulk bicrystalline substrates. Based on vector substrate technology, this novel approach makes use of a few tens-of-nanometers-thick bicrystalline membranes transferred onto conventional substrates. We demonstrate 24° [001]-tilt YBa$_2$Cu$_3$O$_{7-x}$ Josephson junctions fabricated on sapphire single crystals by utilizing 10-nm-thick bicrystalline SrTiO$_3$ membranes.

This technique allows one to manufacture bicrystalline Josephson junctions of high-$T_c$ superconductors on a large variety of bulk substrate materials, providing novel degrees of freedom in designing the junctions and their electronic properties. It furthermore offers the capability to replace the fabrication of bulk bicrystalline substrates with thin-film growth methods.

*v.harbola@fkf.mpg.de

**Introduction**

Bicrystal Josephson junctions of high-transition-temperature (high-$T_c$) cuprate superconductors[1-4] have found many uses in fundamental science and applications.[2, 5-7] For example, bicrystal and the related tricrystal junctions have been instrumental in identifying the *d*-wave order-parameter symmetry of high-$T_c$ cuprates.[8, 9] Studies of their critical currents as a function of the grain boundary angle revealed the necessity of aligning the grains in high-$T_c$ cables and/or utilizing large grain boundary areas[6] to achieve high critical currents, as is now done in coated conductor technologies (see, e.g., Refs.[10, 11]).

Bicrystal Josephson junctions rely on bulk bicrystalline substrates for the epitaxial growth of bicrystalline high-$T_c$ superconducting thin films. However, such substrates are expensive to fabricate, are only available with a limited range of materials and may have properties detrimental to desired applications. Sapphire, for example, features a low dielectric loss tangent of <$10^{-4}$ at $10^{10}$ Hz and 300 K, high thermal conductivity of ≈35 W/mK at 300 K, and a moderate dielectric constant of ≈10. It therefore appears to be a desirable material for substrates of high-$T_c$ Josephson junctions.[12] However, because of its hexagonal crystal lattice symmetry, its poor lattice match, and its contamination of high-$T_c$ films by Al diffusion, sapphire is not generally considered a useful material for bicrystals.

Recently, a new approach called vector substrate technology[13] was presented, that builds on advances based on the fabrication[14-20] and application[21-28] of freestanding oxide membranes and offers new degrees of freedom in choosing substrates for the epitaxial growth of thin films.[13, 28] With vector substrate technology, membranes as thin as ten nanometers placed on carrier substrates are used as templates for the epitaxial film growth of further films. Importantly, only the membrane has to have the properties required for further epitaxial film growth, whereas the materials of the carriers are constrained by far fewer restrictions. For example, a carrier may be an unrefined low-cost compound or a material with properties that offer advantages for a desired application. This technology is versatile and allows epitaxial growth of single-crystalline films on curved surfaces, among many other uses.

Given its many advantages, vector substrate technology has been proposed as an alternative to bicrystalline bulk substrates by making use of bicrystalline membranes[13] as illustrated in Figure 1a. Here, we report on the successful realization of this concept by demonstrating excellent bicrystalline YBa$_2$Cu$_3$O$_{7-x}$ Josephson junctions fabricated on standard *c*-plane sapphire single crystals. In this demonstration, the surfaces of the sapphire substrates are coated by 10-nm-thick SrTiO$_3$ membranes. These membranes are obtained by thin-film growth on bicrystalline substrates, called parent substrates, which can be recycled for growing numerous additional membranes.

To fabricate bicrystalline YBa$_2$Cu$_3$O$_{7-x}$ Josephson junctions on sapphire single crystals, we used SrTiO$_3$ bicrystals[29] as parent substrates. On such a parent substrate, a water-soluble, sacrificial Sr$_2$CaAl$_2$O$_6$ layer[23, 30] and a SrTiO$_3$ layer were epitaxially deposited by pulsed laser deposition (PLD), such that the bicrystal structure was transferred through the sacrificial layer to the final SrTiO$_3$ membrane. By dissolving the sacrificial layer in water, the bicrystal SrTiO$_3$ membrane was then lifted off and transferred onto a carrier substrate, in our case a sapphire single crystal, yielding a vector substrate. In this process, illustrated in Figure 1b, the surface structure of the bicrystal parent substrate remains unscathed. The parent substrates are therefore reusable, such that numerous membranes can be derived from one parent substrate (Figure S2). Furthermore, vector substrates fabricated in this way may also be used as parent substrates, obviating the need for bulk bicrystals completely.

In detail, the samples were prepared by first terminating 24° [001]-tilt SrTiO$_3$ bicrystals[29] with a TiO$_2$ plane by an *in situ* anneal in $pO_2$= 0.1 mbar at 825°C for 40 min such that their surfaces exhibited standard step-terrace patterns and well visible boundaries between the two constituent crystals. Following the layer-by-layer growth of Sr$_2$CaAl$_2$O$_6$—this material having been selected for its lattice match to SrTiO$_3$—by PLD, SrTiO$_3$ layers were also grown layer-by-layer by PLD (Supporting Information Figure S1, see Supporting Information). To provide structural support for the subsequent transfer of the membranes, these samples were spin-coated with ~600-nm-thick PMMA layers. Upon dissolving the sacrificial layers in water, the PMMA-covered bicrystalline SrTiO$_3$ membranes were transferred in air onto thermally prepared sapphire substrates (Figure 1b, see Supporting Information).

The surfaces of the parent substrates were restored by cleaning and annealing to again reveal the distinct step-terrace features (Supporting Information Figure S2). We reused the parent substrates up to five times without encountering the end of their lifetime. Note, however, that although the surfaces of the parent bicrystal substrate can be completely restored, the bicrystal grain boundaries may experience some reconstruction due to the high temperatures used for film growth and annealing, resulting in a sub-nanometer trench that develops along the boundary during subsequent preparations (Supporting Information Figure S2).

Whereas the fragile grain boundary sections of the membranes remained intact after the transfer, the outer parts of the membranes always became fragmented. As a result, the bicrystal membranes had variable shapes with areas of several mm$^2$ (Supporting Information Figure S4).

To prepare the vector substrate surface and remove contaminants still present between the sapphire substrates and the SrTiO$_3$ membranes and to enhance the bonding between the membranes and the sapphires, a 825°C, 40-min anneal in $pO_2$= 0.1 mbar was also performed on these samples.[13, 31] During this anneal, the sample surfaces smoothly bulged upward by ~20 nm over a length perpendicular to the boundary of ~2×200 nm$^2$, which we attribute to relaxation of strain between the two SrTiO$_3$ grains. Otherwise we found the surfaces to be flat and smooth, well suited for subsequent epitaxial growth. Figure 2 provides a series of atomic force (AFM) micrographs that illustrate the evolution of the sample surface during its fabrication.

These results reveal that it is possible to grow heterostructures consisting of a bicrystalline substrate and several layers of thin films, in which the grain boundary is copied from the substrate into each layer. It is even possible to do so in case one of the layers, in the present case the Sr$_2$CaAl$_2$O$_6$ film, has a non-perovskite structure. Furthermore, the results show that bicrystalline membranes are sufficiently robust to allow them to be lifted off and transferred.

In the next step, ~100-nm-thick YBa$_2$Cu$_3$O$_{7-x}$ films were epitaxially grown by PLD on the bicrystalline SrTiO$_3$ surface using standard growth parameters (see Supporting Information). The film surfaces were decorated by precipitates (Figure 2e) attributed to a non-stoichiometric material transfer from the PLD target to the growing film. X-ray diffraction revealed the films to be otherwise single-phase and $c$-axis oriented consistent with YBa$_2$Cu$_3$O$_{7-x}$ films grown on standard SrTiO$_3$ single-crystalline substrates.[32] Their intragrain superconducting $T_{c,0}$ equaled ~88 K. To define ~3–4-$\mu$m-wide superconducting bridges crossing the grain boundary as desired to fabricate the Josephson junctions, the samples were patterned by optical lithography. The patterning called for the deposition of an ~20-nm-thick Au film on the sample as well as that lithography, gold etching and Ar-ion milling be performed. The samples were resilient through each stage of the fabrication process. A scanning electron microscopy image of a typical sample is shown in Figure 2f.

The transport properties of the bridges were obtained from 4-point measurements performed at 4.2 K in an electrically and magnetically shielded environment with an external magnetic field applied perpendicular to the substrate plane. Figure 3 provides the current–voltage $I(V)$ characteristics for two nominally identical bridges under the influence of the magnetic field. Both bridges show textbook-like $I(V)$ characteristics, as described by the resistively and capacitively shunted junction (RCSJ) model.[33, 34] The critical currents $I_c$ at $H = 0$ equal ~116 and 159 µA, corresponding to critical current densities $J_c$ ~$5\times10^4$ A/cm$^2$, respectively. These $I_c$ values were obtained after keeping the samples at 300 K for several months. Immediately after fabrication, they were a factor of about 3–5 smaller (Supporting Information Figure S6). We attribute this enhancement to relaxation of oxygen ions at the grain boundary.[2] The normal state resistance $R_n$ of the junctions shunted by the 18-nm-thick Au layer were calculated to equal 7.1 and 5.4 Ω. The calculated $I_cR_n$ products for the two bridges are 820 and 850 µV, in agreement with literature values of ~300 µV of comparably shunted 24° junctions.[35, 36] The fit of the $I(V)$ characteristics by the RCSJ model yields a Stewart–McCumber parameter of $\beta_c$ ~1.7, which agrees with the expected value of $\beta_c$ ~2.5 resulting from a stray capacity of ~0.15 pF caused by the huge low-temperature dielectric permittivity of SrTiO$_3$.[37] The $I_c(H)$ characteristics (Figure 4) reveal a notably homogeneous current density and high-quality junctions.

In summary, we have demonstrated the viability of fabricating bicrystalline high-$T_c$ Josephson junctions on bulk single-crystalline sapphire substrates by using vector substrate technology. This process utilizes membranes of bicrystalline films, which are deposited by thin-film growth methods and transferred onto substrates of choice. We have demonstrated the viability of this technique by fabricating high-quality, 24° YBa$_2$Cu$_3$O$_{7-x}$ bicrystalline Josephson junctions. This technique eliminates the need for bicrystalline substrates to grow bicrystal Josephson junctions, and it allows the growth of such junctions on bulk substrates of choice, which are only covered by a ~10-nm-thick bicrystalline layer suitable for epitaxial growth. This technology can be readily extended (i) to fabricate further vector substrates, such as tricrystalline or polycrystalline substrates, (ii) to grow bicrystalline films of other, possibly non-superconducting materials, and (iii) to deposit heterostructures that comprise single-crystalline films on top of bicrystal layers.

**Acknowledgments**

This project was partially supported by EU FLAG-ERA Project To2Dox, the European Commission under H2020 FET Open Grant ''FIBsuperProbes'' (Grant No. 892427), the COST action SUPERQUMAP (CA2114), and the German Science Foundation (DFG). We thank L. Pavka for editorial support.

**Figures:**

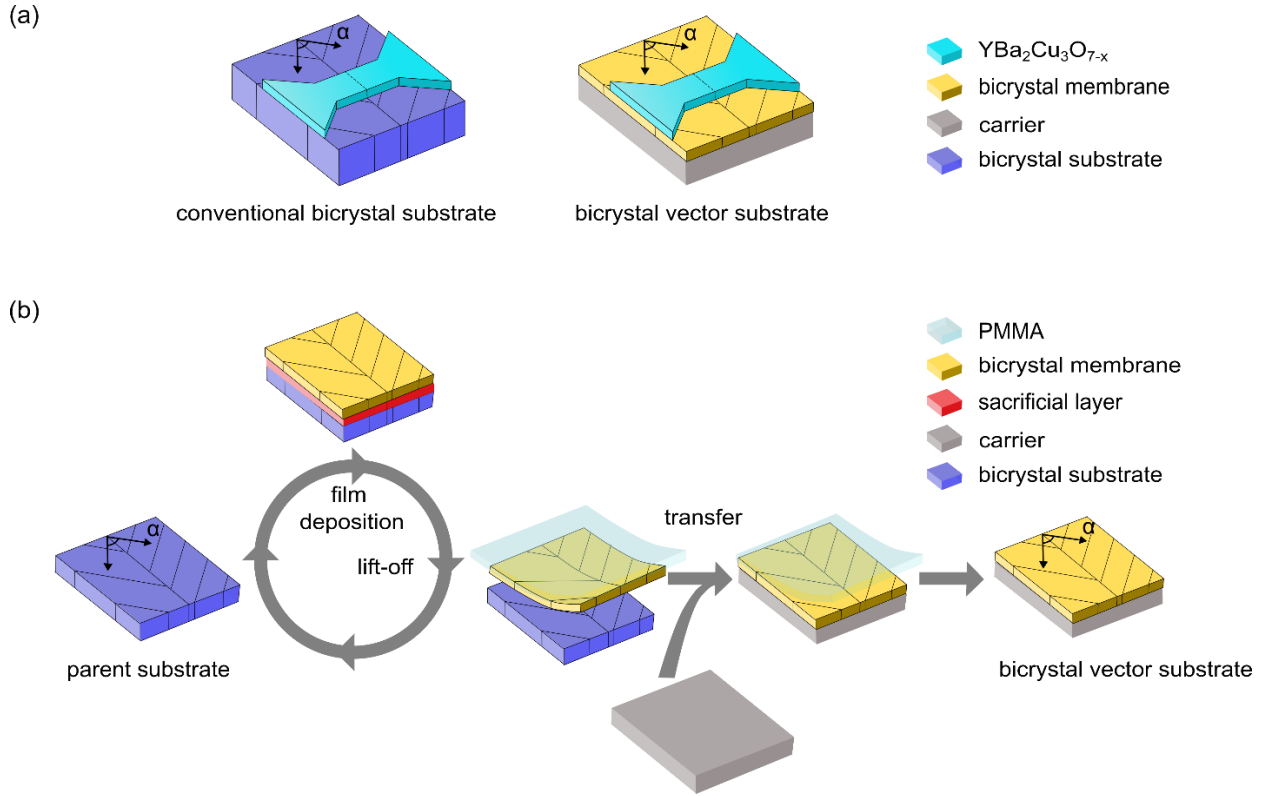

**Figure 1.** (a) Illustration of the main idea of this work. Left: sketch of an arrangement of a bicrystalline Josephson junction fabricated by epitaxial growth on a conventional bicrystalline bulk substrate. Right: sketch of a bicrystal arrangement that utilizes a vector substrate. The vector substrate, which consists of a bulk substrate of almost unrestricted choice, carries a bicrystalline thin-film membrane sintered to its surface. In this case, it is a sapphire single crystal and a 10-nm-thick, 24° SrTiO$_3$ bicrystalline film. High-$T_\mathrm{c}$ superconducting films grown on either substrate yield Josephson junctions at the bicrystal grain boundary. (b) Fabrication process of the vector substrates. Note that the parent substrates can be recycled many times. Here, α equals ~24º.

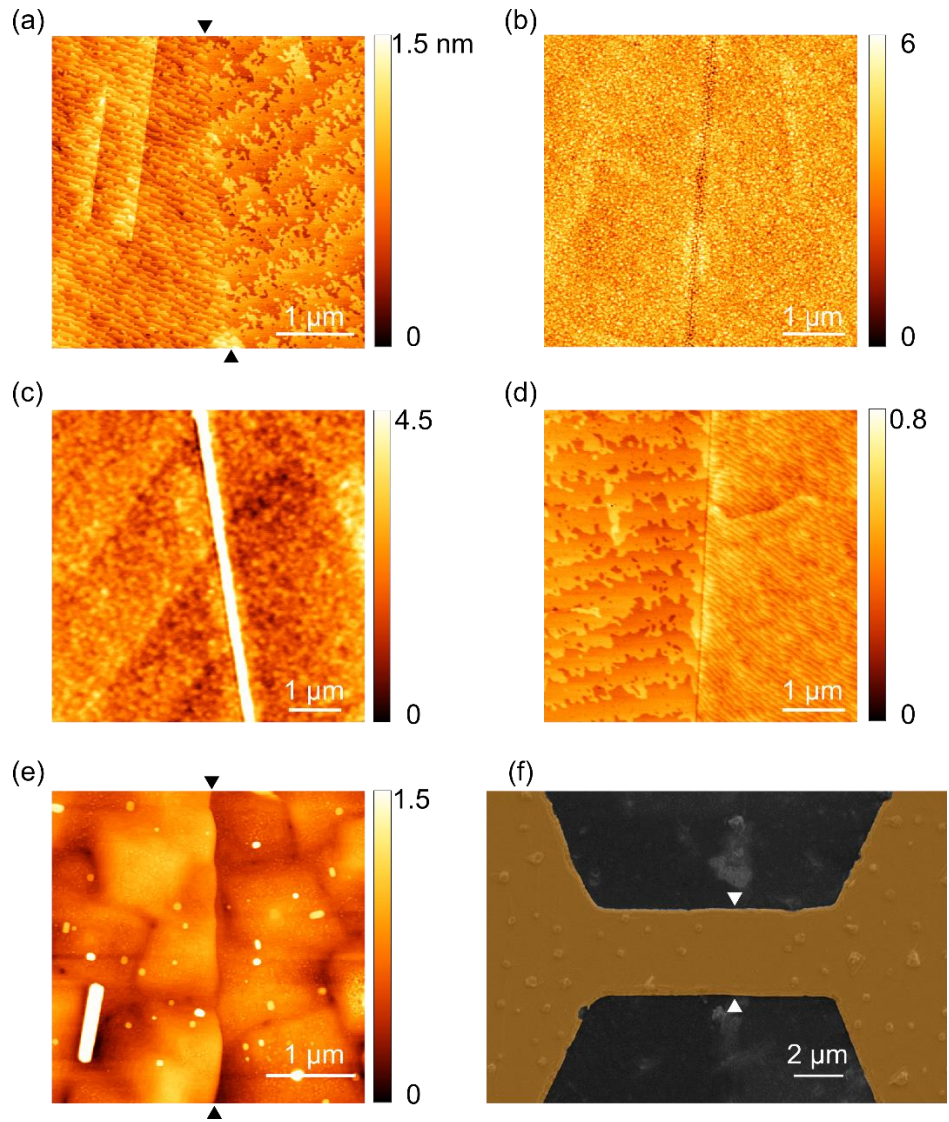

**Figure 2.** (a)–(e) AFM micrographs of sample surfaces imaged during various stages of sample preparation. (a) Surface of a bicrystal parent substrate after termination at 825°C for 40 min. (b) Surface morphology after the growth of $Sr_2CaAl_2O_6$ layers with a $SrTiO_3$ membrane at top. (c) The same bicrystal $SrTiO_3$ membrane after transfer onto a $c$-plane sapphire and post-annealing at 825°C for 40 min. (d) Surface morphology of the parent bicrystal substrate after membrane lift-off, cleaning and preparation for reuse. (e) Grain boundary after deposition of $YBa_2Cu_3O_{7-x}$ on a bicrystal vector substrate. (f) Scanning electron microscopy image of a $YBa_2Cu_3O_{7-x}$ microbridge crossing the 24° [001]-tilt bicrystal boundary of the vector substrate, which forms a Josephson junction. Grain boundary positions in (e) and (f) are marked by arrows.

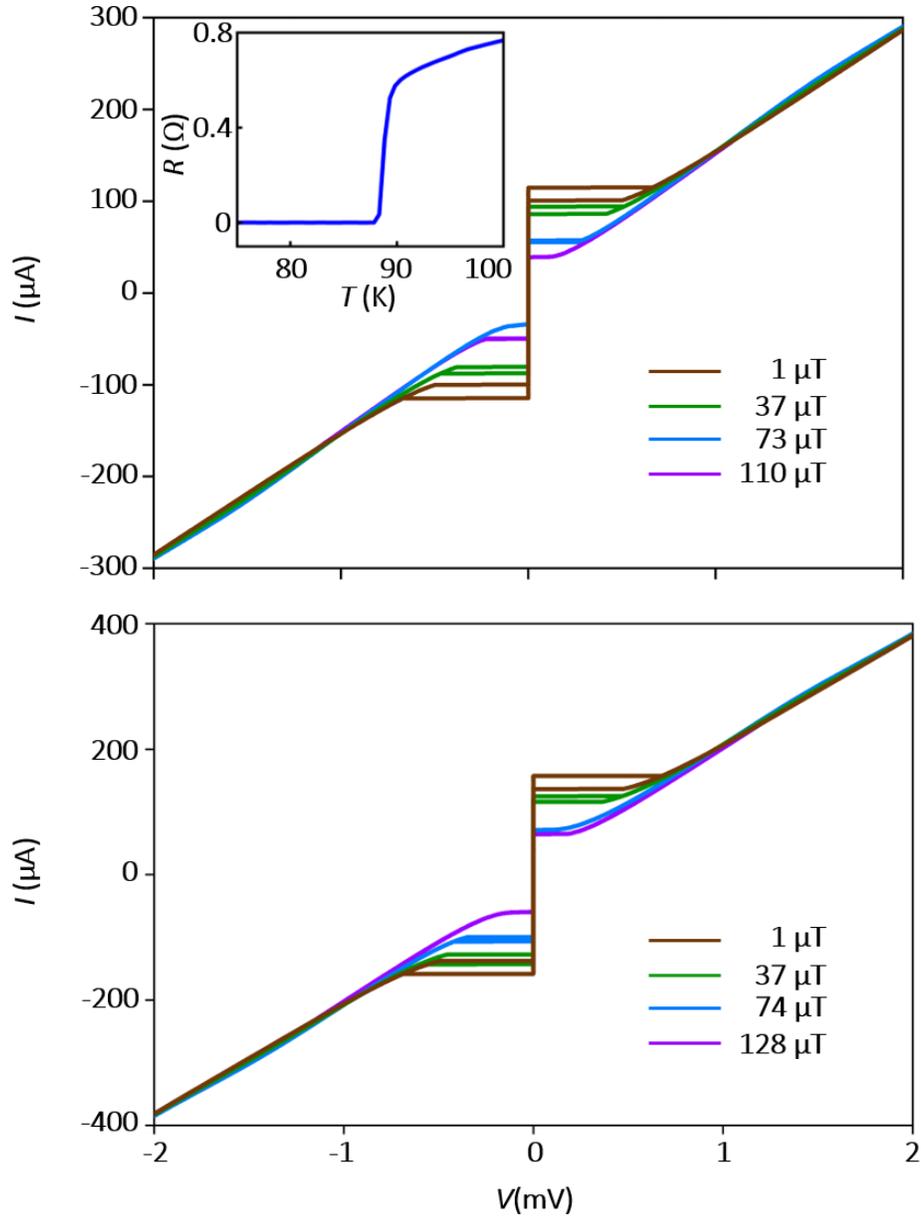

**Figure 3.** Current–voltage characteristics of two nominally identical 24° bicrystal Josephson junctions fabricated with vector substrate technology. The data were taken at 4.2 K with a magnetic field $H$ applied perpendicular to the sample surfaces as described. The characteristics follow RCSJ behavior (see Supporting Information Figure S8). Inset in the upper graph: Intragrain $R(T)$ characteristic of a nominally identical sister sample.

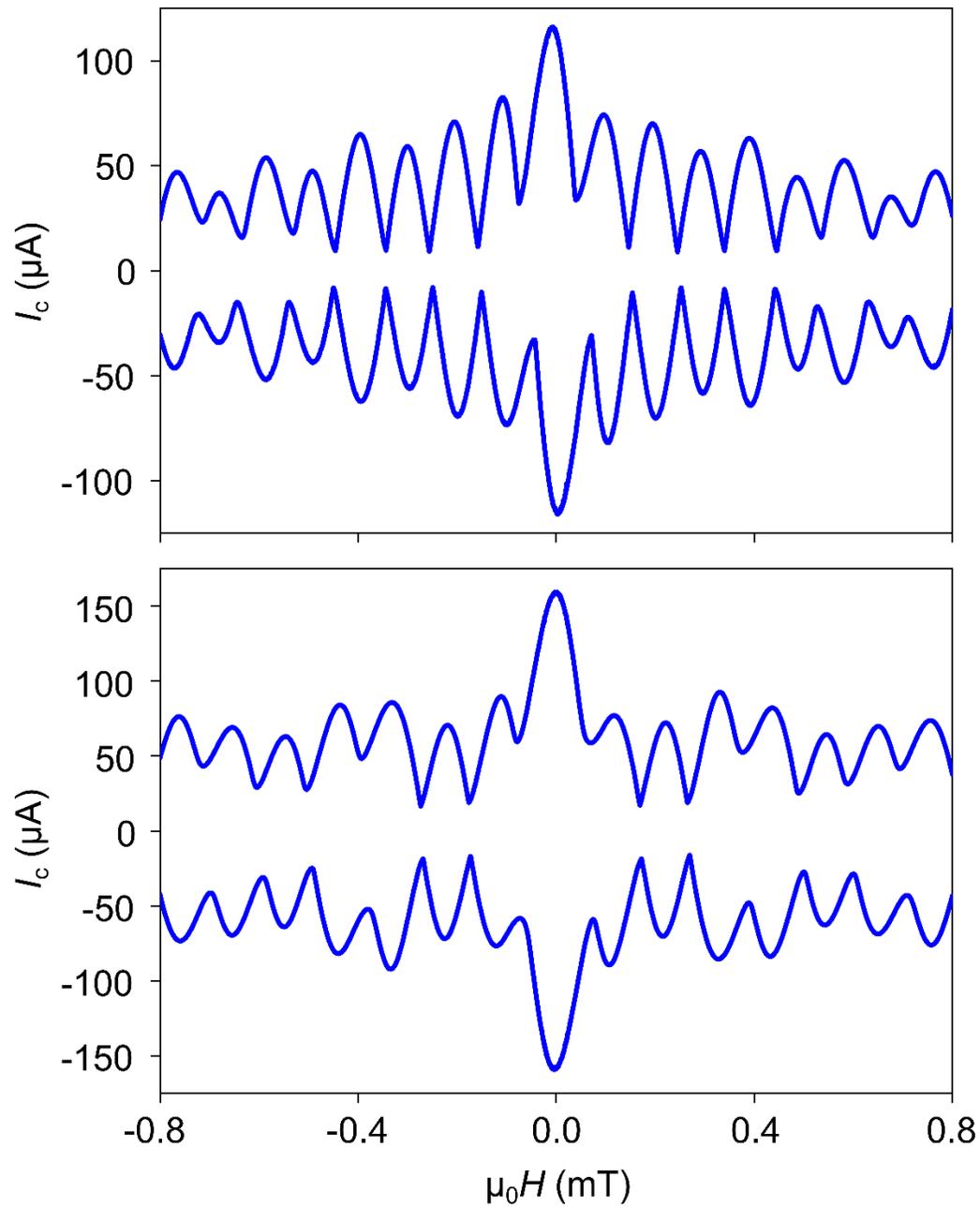

**Figure 4.** $I_c(H)$ characteristics of the two Josephson junctions shown in Figure 3 measured at 4.2 K. The magnetic field $H$ is applied perpendicular to the sample surface.

Supporting Information for

**Vector-substrate-based Josephson junctions**

Yu-Jung Wu[1], Martin Hack[2], Katja Wurster[2], Simon Koch[2], Reinhold Kleiner[2], Dieter Koelle[2], Jochen Mannhart[1] & Varun Harbola[1]*

[1]Max Planck Institute for Solid State Research

Heisenbergstrasse 1, 70569 Stuttgart, Germany

[2]Physikalisches Institut, Center for Quantum Science (CQ) and LISA+

Universität Tübingen, auf der Morgenstelle 14, 72076 Tübingen, Germany

*v.harbola@fkf.mpg.de

**Supporting note S1:**

*Epitaxial Film Growth:* Prior to growth, the SrTiO$_3$ bicrystal substrates were pre-annealed in $pO_2$= 0.1 mbar at 825°C for 40 minutes using a CO$_2$ laser to attain a smooth step-terrace surface. Subsequently, the Sr$_2$CaAl$_2$O$_6$ sacrificial layer and SrTiO$_3$ top membrane were deposited via pulsed laser deposition, utilizing an excimer laser with 248 nm wavelength. The Sr$_2$CaAl$_2$O$_6$ sacrificial layer was deposited at a substrate temperature of 825°C and $pO_2$= 1 x 10$^{-5}$ mbar, with a laser fluence of 1.76 Jcm$^{-2}$. The SrTiO$_3$ membrane was grown at a substrate temperature of 825°C and $pO_2$= 1 x 10$^{-5}$ mbar, with a laser fluence of 1.21 Jcm$^{-2}$. The films were grown in the layer-by-layer mode, with film thickness monitored via RHEED oscillations (see Figure S1). YBa$_2$Cu$_3$O$_{7-x}$ films were grown at 720°C and $pO_2$= 0.25 mbar, 2 J cm$^{-2}$ laser fluence, followed by post-annealing at 450°C and $pO_2$= 450 mbar for 1 hr. All the targets were commercial targets from Kurt J. Lesker.

*Transfer and Characterization of Freestanding Membranes:* PMMA layers were spin-coated onto the SrTiO$_3$- Sr$_2$CaAl$_2$O$_6$- SrTiO$_3$ stacks. The stacks were immersed in room-temperature deionized water to dissolve the Sr$_2$CaAl$_2$O$_6$ sacrificial layers. Following lift-off with PMMA support, the resulting freestanding membranes were affixed onto annealed Al$_2$O$_3$ substrate by heating them to 80°C on a hot plate. The Al$_2$O$_3$ had been prepared by annealing at 1615°C in vacuum to achieve a smooth, well-defined step-and-terrace surface. The membranes remained on the substrate subsequent to the dissolution of the PMMA layers with acetone. All the AFM images were captured using an Asylum Cypher AFM in tapping mode. The X-ray diffraction data were obtained utilizing a monochromatic Cu-K$_{\alpha 1}$ source on a Panalytical Empyrean machine.

*Substrates Reprepare and Reuse:* Following dissolution and lift-off, the bicrystal parent substrate underwent a cleaning process with deionized water, acetone, and isopropanol in a ultrasonic bath for 10 minutes each. After cleaning, the substrates were then annealed at $pO_2$= 7.5 x 10$^{-2}$ mbar at 825°C for 8 minutes. As a result of the treatment, the surface of the reused substrate then regained cleanliness and exhibited clear step-terrace surface.

*Josephson Junction Fabrication and Measurement:* 18 nm of gold (Au) were thermally evaporated onto YBa$_2$Cu$_3$O$_{7-x}$ at room temperature, serving as a resistive shunt and protection layer during the fabrication of the bridges. The both 3–4 µm wide bridges straddling the grain boundary were fabricated via photolithography and Ar ion milling. The bridges were characterized at $T$ = 4.2 K in an electrically and magnetically shielded environment. For the measurement of current-voltage characteristics, $I_c(H)$ and $V(H)$, a four-point arrangement with a room temperature voltage amplifier was employed.

**Supplementary figures:**

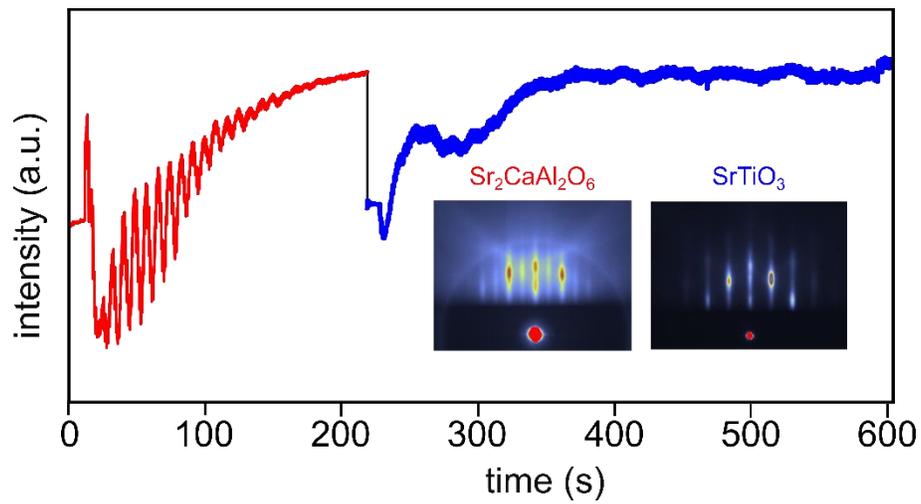

**Figure S1.** Layer-by-layer growth of bicrystal substrate, aluminate sacrificial layer and $SrTiO_3$ heterostructure monitored via *in-situ* RHEED.

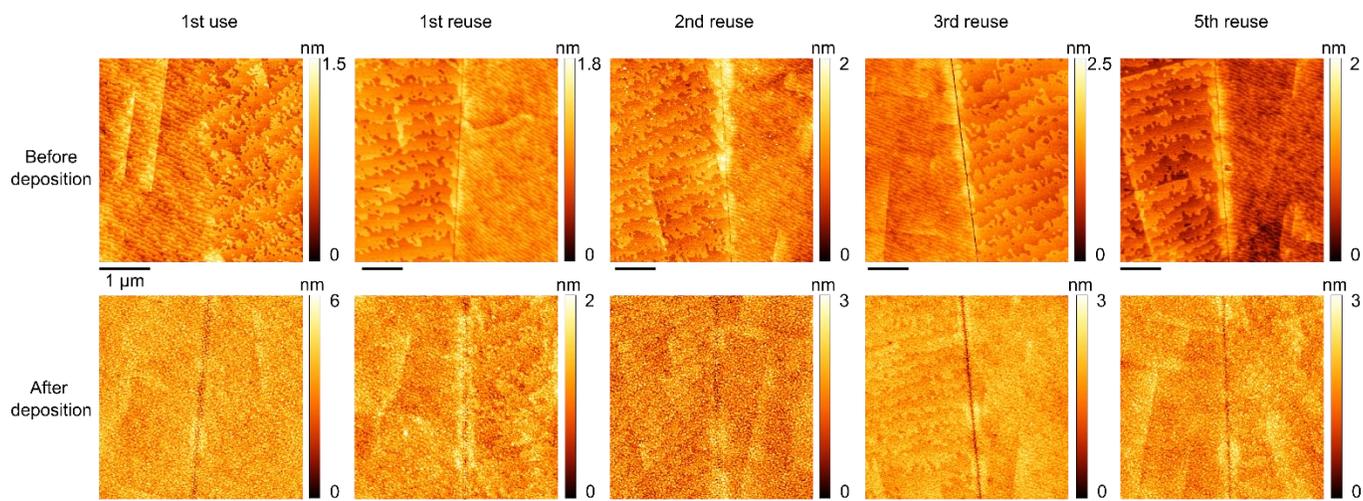

**Figure S2.** Evolution of the surface morphology upon reusing the parent SrTiO$_3$ bicrystal substrate. A clear preservation of the bicrystal grain boundary upon substrate reusage after membrane lift off can be seen here. There is a slightly more pronounced diffusion of material near and at the grain boundary resulting in a sub-nanometer trench appearing even after the first reuse and evolving with subsequent reusage (top row). Bottom row shows how the surface looks after deposition of the sacrificial layer and thin film on top of the parent substrate. Note that even after the 5$^{th}$ reuse, the substrate surface is still viable for epitaxial growth.

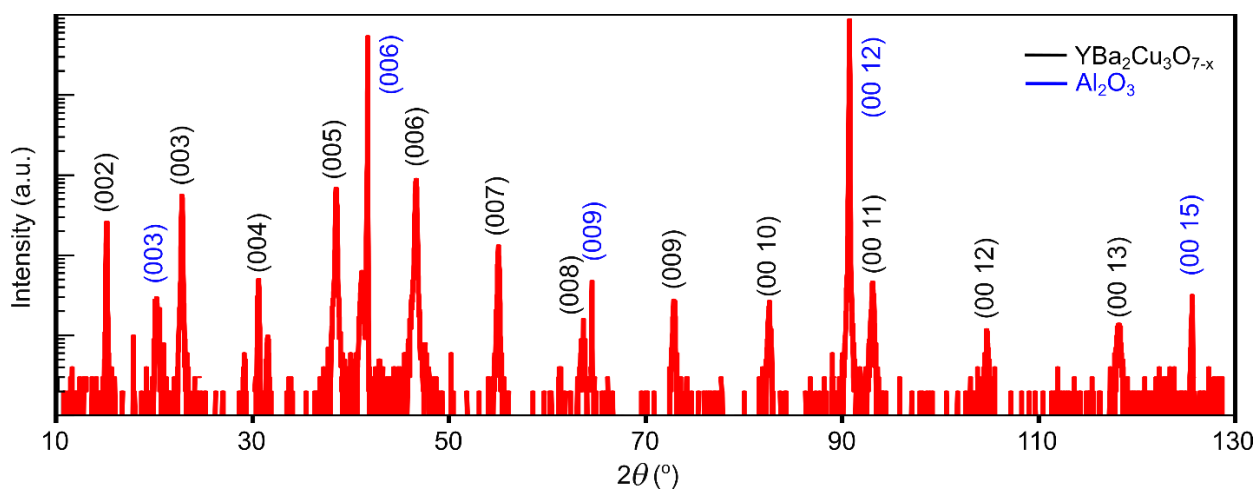

**Figure S3.** A 2$\theta$-$\omega$ X-ray diffraction scan of YBa$_2$Cu$_3$O$_{7-x}$ deposited on a bicrystal SrTiO$_3$-on-sapphire vector substrate. A single-phase growth of YBa$_2$Cu$_3$O$_{7-x}$ occurs with peaks corresponding to YBa$_2$Cu$_3$O$_{7-x}$ on SrTiO$_3$ (001) identified explicitly here. The SrTiO$_3$ peaks are not perceptible here owing to the 10 nm thickness of this layer, compared to ~100 nm of the YBa$_2$Cu$_3$O$_{7-x}$ layer.

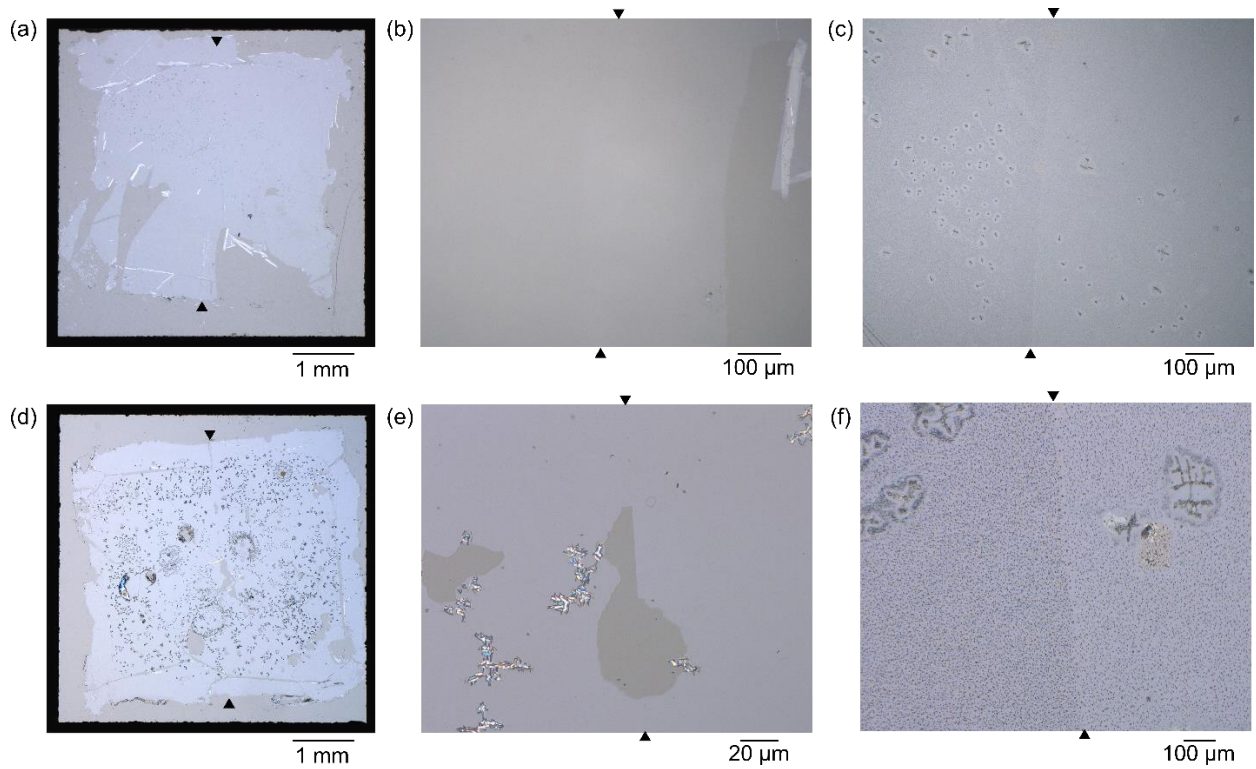

**Figure S4.** Optical microscopy images of two different bicrystal SrTiO$_3$ on sapphire vector substrates made from the same parent substrate. (a) and (d) are images of the full vector substrates on 5x5 mm$^2$ sapphire carriers. (b) and (e) are images zooming into the grain boundary area of the vector substrates, where the grain boundary is optically not visible, but in (e), the fracture of the membrane can be seen to happen right at the grain boundary, indicating its position. By comparison of (d) and (e), what looks like dirt in (d) is actually wrinkling of the bicrystal membrane upon this transfer, as apparent in (e). (c) and (f) are images taken after deposition of ~ 100 nm YBa$_2$Cu$_3$O$_{7-x}$ on these vector substrates, and now the bicrystal boundary is perceptibly visible. The arrows mark the position of the grain boundaries in all the images.

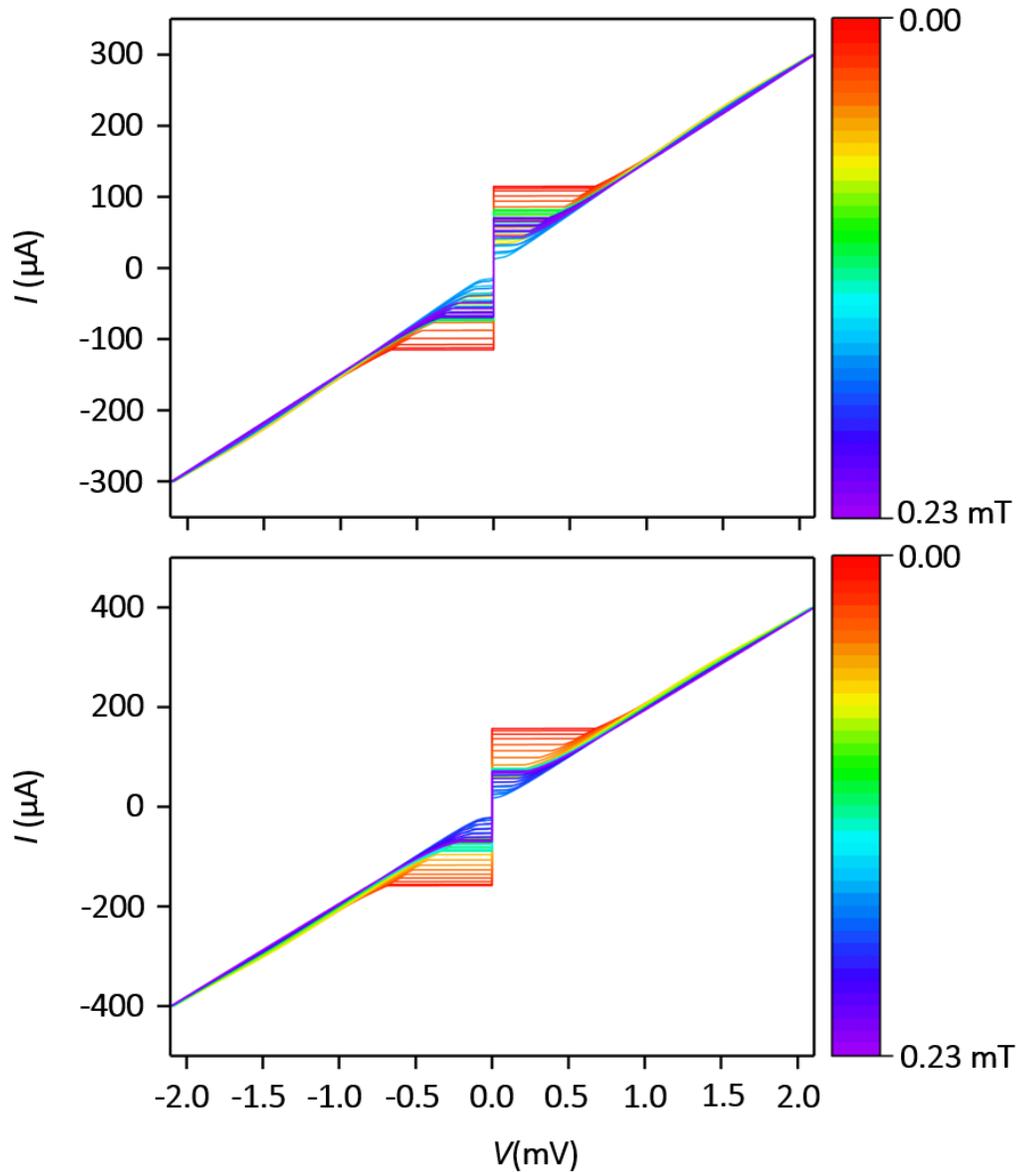

**Figure S5.** The complete and extended dataset of the Josephson junction characteristics shown in Figure. 3 of the main text. Electric-transport characteristics of YBa$_2$Cu$_3$O$_{7-x}$ grain boundary Josephson Junctions fabricated with the vector substrate technology. Shown are several Current vs Voltage measurements at different external fields ranging from 0 to 0.23 mT. Only one direction of the current sweep is shown here for clarity.

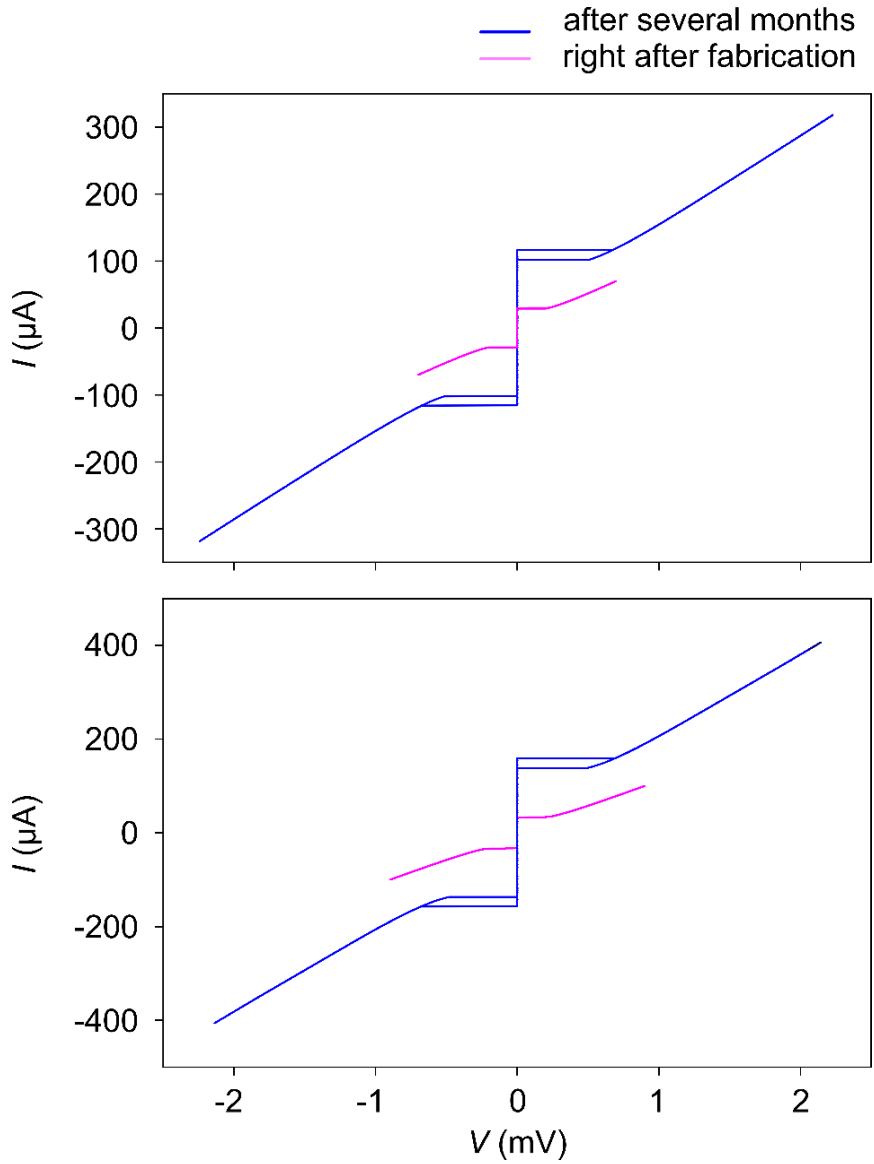

**Figure S6.** Temporal evolution of the *I*(*V*) characteristics of two Josephson junctions shown in Figure 3. Both junctions were measured right after fabricating them and after several months of storage in a nitrogen atmosphere. An improvement in the $I_c$ is evident, by a factor of about five in both junctions characterized, along with a more pronounced hysteresis during the *I-V* sweep, owing to the increased Stewart-McCumber parameter. This "healing" of the Josephson junction can be attributed to the oxygen ion relaxation at the grain boundary with time.

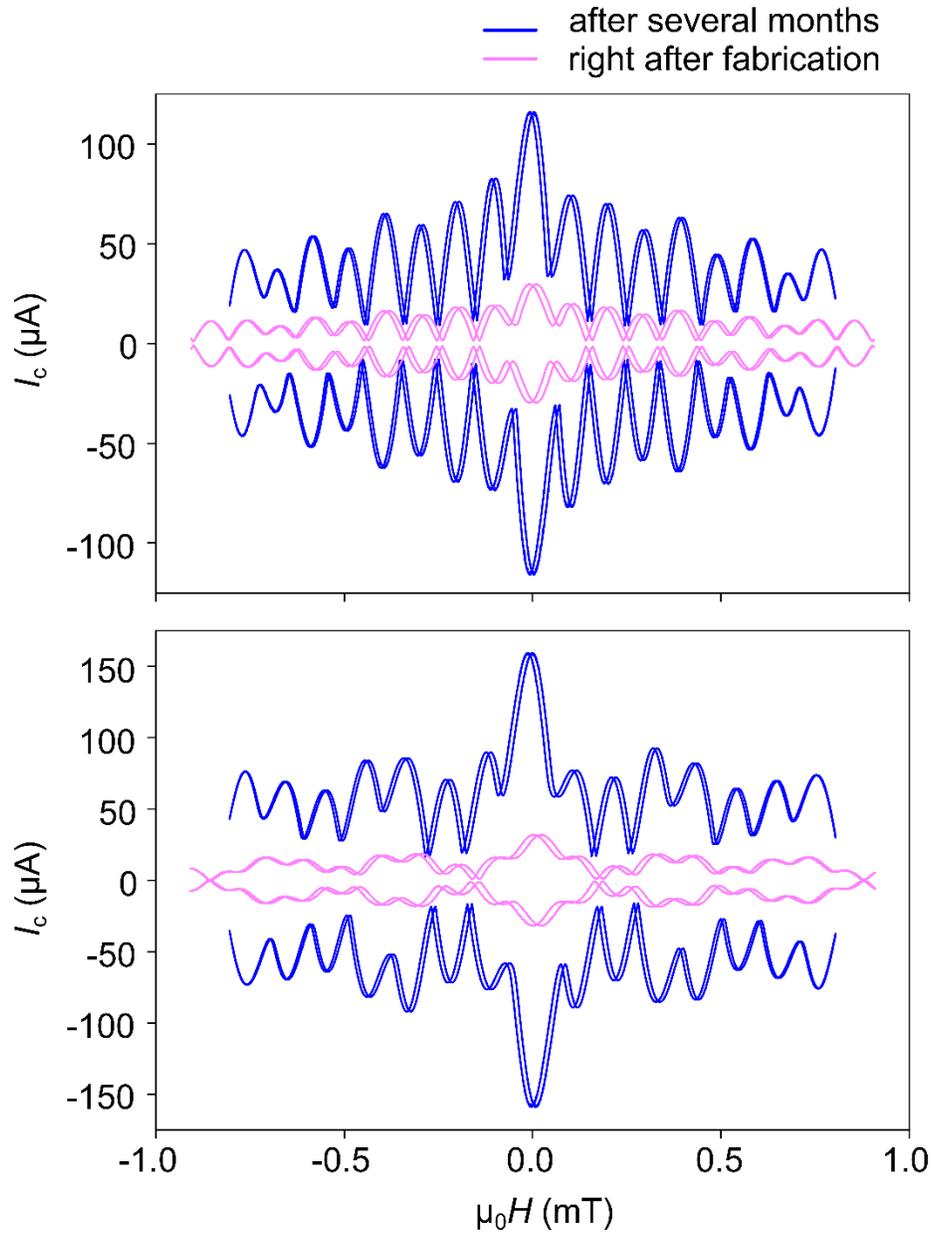

**Figure S7.** Temporal evolution of $I_c(H)$ of the same two Josephson junctions as in Figure S6. The shape of the $I_c$ oscillations is retained over time but an overall increase in $I_c$ by about a factor of five applies also for the small magnetic fields under which the Josephson junctions were characterized after a span of several months.

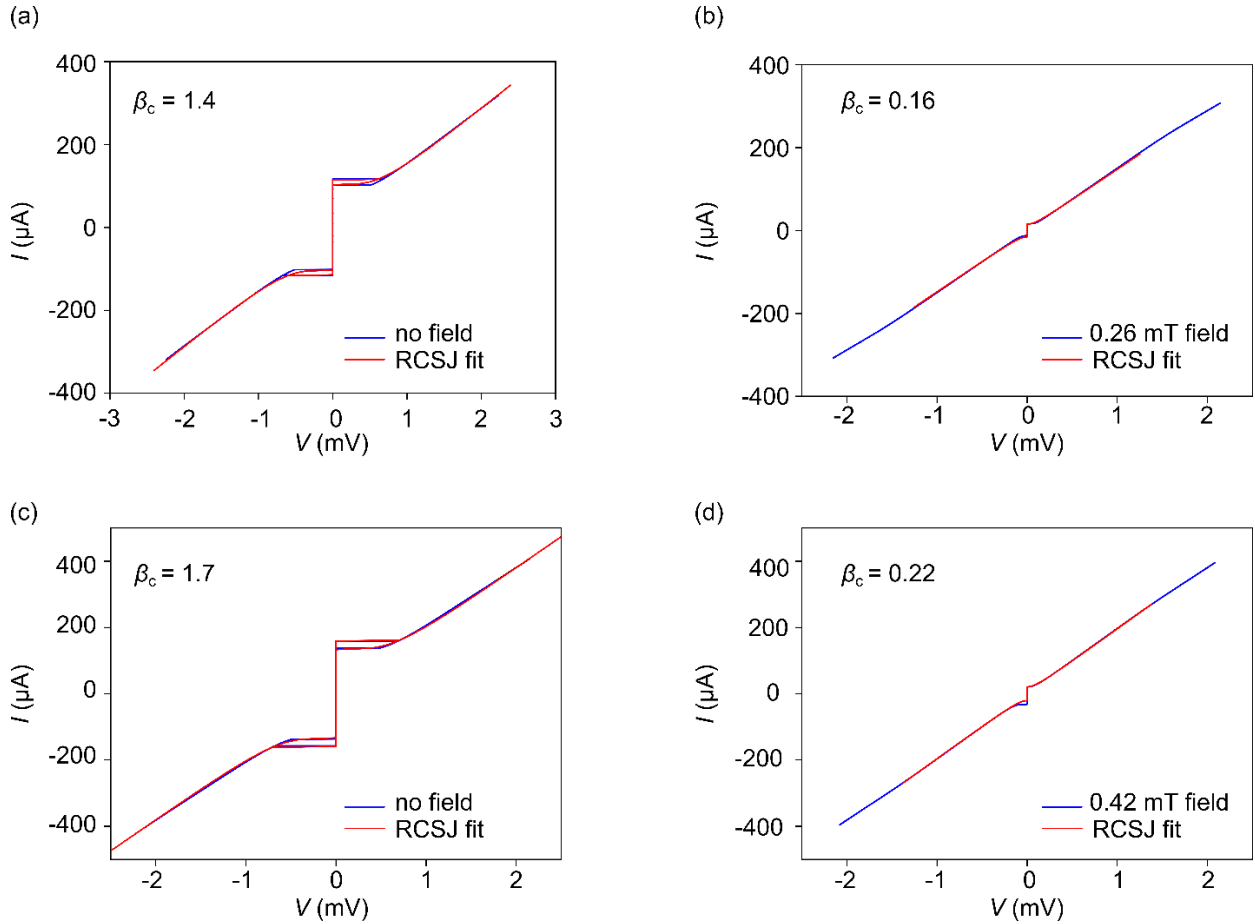

**Figure S8.** Electric-transport characteristics of the two Josephson junctions shown in Figure 3. Shown are current (*I*) vs voltage (*V*) measurements compared to RCSJ fits to determine $β_c$ of the Josephson junctions. (a), (b) *I-V* characteristics of the bridge shown in Figure 3a at 0 and 0.26 mT external field, repectively. The comparison with the fit results in a $β_c$ of 1.4 and 0.16, respectively. (c), (d) *I-V* characteristics of the other bridge shown in Figure 3b at 0 and 0.42 mT external field, respectively. The comparison with the fit results in $β_c$ of 1.7 and 0.22, respectively.